\documentclass[showpacs,preprintnumbers,amsmath,amssymb]{revtex4}

\usepackage{graphicx}
\usepackage{dcolumn}
\usepackage{bm}

\begin{document}

\title{Nonequilibrium steady state transport of collective-qubit system in strong coupling regime}

\author{Chen Wang$^{1}$}\email{wangchen@hdu.edu.cn}
\author{Kewei Sun$^{1}$}

\address{
$^{1}$Department of Physics, Hangzhou Dianzi University, Hangzhou, Zhejiang 310018, China
}

\date{\today}

\begin{abstract}
We investigate the steady state photon transport in a nonequilibrium collective-qubit model.
By adopting the noninteracting blip approximation, which is applicable in the strong photon-qubit coupling regime,
we describe the essential contribution of indirect qubit-qubit interaction to the population distribution,
mediated by the photonic baths.
The linear relations of both the optimal flux and noise power with the qubits system size are obtained.
Moreover, the inversed power-law style for the finite-size scaling  of the optimal photon-qubit coupling strength is exhibited,
which is proposed to be universal.
\end{abstract}



\maketitle

\section{Introduction}
Deep understanding and optimal control of quantum propagation in low-dimensional
light-qubit (atom) hybrid systems is of fundamental interest and practical importance~\cite{awallraff2004nature1}.
Particularly, the collective effect on the information and energy transport due to light-qubit (atom) scattering, as a novel measuring feature, has recently attracted dramatic attention~\cite{tbrandes2005physrep1,lzhou2008prl1,dzxu2013pra1,dwwang2015prl1}.
Many works have been carried out to observe such effect, ranging from solid state physics~\cite{jklinder2015pnas1,mfeng2015nc1}, quantum biology~\cite{mmohseni1}, to quantum optics~\cite{hritsch2013rmp1,krahazzard2014prl1}.
The transport scheme, quantum flow from hot source to cold drain, can be typically  established by applying the thermodynamic (e.g., temperature) bias, in accordance with the second law of thermodynamics.

The prototype paradigm of collective-qubit systems to characterize quantum transport based on photon-qubit interaction, is termed as
Dicke model. It was originally pointed out by R. H. Dicke,
which described $N$ identical two-level atoms coupled to single radiation field mode~\cite{rhdicke1954pr1}.
Dicke model has been extensively studied in quantum superradiant phase transition associated with collective self-organization of atoms,
which predicts the universal finite-size scaling effect~\cite{nlambert2004prl1,qhchen2008pra1}.
Moreover, the influence of photon dissipation on the collective-qubit model is analyzed by considering continuous radiation modes.
The anomalous superradiant-like relaxation, dynamical quantum beat and quantum phase transition have been unraveled, which significantly
differ from the counterpart in spin-boson model ($N=1$)~\cite{tvorrath2004cp1,tvorrath2004prl1,fbanders2008njp1}.

Recently, steady state transport behavior of collective qubits weakly coupled to two photonic baths is studied together within Redfield scheme
and full counting statistics~\cite{mesposito2009rmp1,mcampisi2011rmp1}, where the thermodynamic bias is applied to drive the unidirectional photonic flow~\cite{mvogl2011aop1,mvogl2012pra1}.
The influence of bath temperatures on the generation of collective quantum transport is intensively analyzed.
As is known in nonequilibrium spin-boson model, strong system-bath coupling plays nontrivial role to exhibit nonmonotonic flux behaviors, by including the nonadditive and nonresonant transport processes between qubits and bosonic (e.g., phonon) baths~\cite{ajleggett1987rmp1,dsegal2006prb1,kavelizhanin2010jcp1,tchen2013prb1,cwang2015sr1}.
Hence, it is indeed desirable to study the effect of strong photon-qubit interaction on quantum transport in collective-qubit system.

In the present work, we adopt the nonequilibrium noninteracting-blip approximation (NIBA)~\cite{ajleggett1987rmp1,hdekker1987pra1,lnicolin2011prb1,lnicolin2011jcp2,tchen2013prb1} to study the steady state collective transport behaviors of multi-qubits system in strong coupling regime.
The indirect qubit-qubit interaction mediated by photonic baths, is clearly revealed, contributing to collective behaviors.
Then, the effect of strong photon-qubit coupling on photonic energy current fluctuations (e.g., flux, noise power) is investigated,
and the finite-size scaling of the corresponding optimal observables is discussed.
This paper is organized as follows: in section II, we describe the collective-qubit model and derive the quantum kinetic equation
in collective-angular momentum basis. The comparison of steady state population in strong coupling regime is made with the counterpart
within Redfield scheme.
In section III, the full counting statistics of photonic energy flux is established, and the finite-size scaling of the current fluctuations
is analyzed.
A concise summary is given in the final section.

\begin{figure}[tbp]
\begin{center}
\includegraphics[scale=0.4]{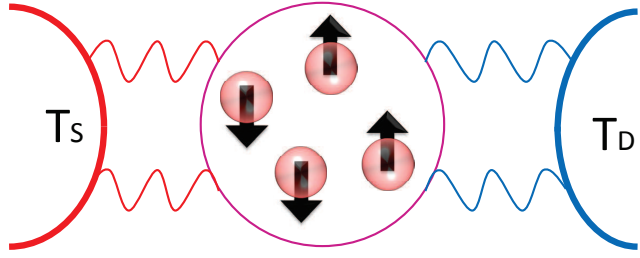}
\end{center}
\caption{(Color online) Schematic description of nonequilibrium qubits system: the left red arch and right blue arch represent the source and drain photonic baths, with the temperature given by $T_S$ and $T_D$, respectively;
the central pink circles combined with the arrows describe the two-level qubits; the cured red and blue lines describes the
system-bath interaction.
}~\label{fig:fig0}
\end{figure}
\section{nonequilibrium collective-qubit system}

\subsection{Model}
The nonequilibrium collective-qubit model at Fig.~\ref{fig:fig0}, composed by $N$ identical two-level qubits interacting with two photonic baths,
is described by
\begin{eqnarray}~\label{ham1}
\hat{H}=-\frac{\epsilon_0}{2}\sum^{N}_{n=1}\hat{\sigma}^{n}_{z}+\frac{\Delta}{2}\sum^{N}_{n=1}\hat{\sigma}^{n}_{x}
+\sum^{N}_{n=1}\sum_{v,k}\lambda^{n}_{v,k}\hat{\sigma}^{n}_{z}(\hat{b}^{\dag}_{k,v}+\hat{b}_{k,v})
+\sum_{k,v}\omega_k\hat{b}^{\dag}_{k,v}\hat{b}_{k,v},
\end{eqnarray}
where $\hat{\sigma}^n_z$ and $\hat{\sigma}^n_{x}$ are the Pauli matrix representing the $n$th two-level qubit,
$\epsilon_0$ is the Zeeman splitting energy of the qubit, $\Delta$ denotes the coherent tunneling strength between two levels of the qubit.
$\hat{b}^{\dag}_{k,v}~(\hat{b}_{k,v})$ creates (annihilates) one photon with the frequency $\omega_k$ and momentum $k$ in the $v$th photonic bath,
and $\lambda_{n,k,v}$ describes the coupling strength between $n$th qubit and photon with momentum $k$ in $v$th bath.
In this paper, we simplify the system-bath coupling independent of the specific qubit $\lambda^{n}_{k,v}=\lambda_{k,v}$.
The interaction of the qubits system with the photonic baths is characterized by the spectral density
$J_v(\omega)=4\pi\sum_k|\lambda_{k,v}|^2\delta(\omega-\omega_k)$.
Here, we assume that the typical Ohmic function characterizes baths~\cite{dsegal2014pre1,ksaito2013prl1,dsegal2015arxiv2}, specified as
$J_v(\omega)=\alpha_v{\omega}e^{-\omega/\omega_{c,v}}$, where
$\alpha_v$ is the coupling between the qubits and $v$th bath, and $\omega_{c,v}$ is the cutoff frequency of the $v$th bath.

For the qubits do not interact with each other, we  introduce the angular momentum operators
$\hat{J}_z=\frac{1}{2}\sum^{N}_{n=1}\hat{\sigma}_z$ and
$\hat{J}_x=\frac{1}{2}\sum^{N}_{n=1}\hat{\sigma}_x$, to describe the collective behaviors of qubit system.
Then, the nonequilibrium collective-qubit model is simplified to
\begin{eqnarray}~\label{ham2}
\hat{H}=-\epsilon_0\hat{J}_{z}+{\Delta}\hat{J}_{x}
+2\hat{J}_{z}\sum_{v,k}\lambda_{v,k}(\hat{b}^{\dag}_{k,v}+\hat{b}_{k,v})
+\sum_{k,v}\omega_k\hat{b}^{\dag}_{k,v}\hat{b}_{k,v},
\end{eqnarray}
where the angular momentum operator $\hat{J}_x=\frac{1}{2}(\hat{J}_{+}+\hat{J}_{-})$, with
$\hat{J}_{\pm}$ creating and annihilating angular momentum. They are described
as $\hat{J}_{\pm}|j,m{\rangle}=\sqrt{g^{\pm}_{m}}|j,m{\rangle}$ under the angular momentum basis $|j,m{\rangle}$,
with $j=N/2$ and $g^{\pm}_{m}=j(j+1)-m(m{\pm}1)$.
They obey such commutating relations
$[\hat{J}_+,\hat{J}_-]=2\hat{J}_z$, and $[\hat{J}_{\pm},\hat{J}_z]={\pm}\hat{J}_{\pm}$.

As two photonic baths have the temperature bias (e.g., $T_S>T_D$), there exists unidirectional photonic flux from hot source to cold drain.
Here, we are interested in the influence of the multi-photon excitations on quantum transport, which usually occurs beyond weak system-bath coupling regime,
it is practically useful to transform the model at Eq.~(\ref{ham2}) by using the canonical transformation to
$\hat{H}_t=\hat{P}^{\dag}\hat{H}\hat{P}$, with the unitary operator $\hat{P}=e^{i\hat{J}_z\hat{B}}$
and the collective photonic momentum operator
\begin{eqnarray}
\hat{B}=\sum_{v=S,D}\hat{B}_v=i\sum_{k,v}\frac{2\lambda_{k,v}}{\omega_k}(\hat{b}^{\dag}_{k,v}-\hat{b}_{k,v}).
\end{eqnarray}
The transformed Hamiltonian is shown as
$\hat{H}_t=\hat{H}_s+\hat{V}_{sb}+\sum_{k,v}\omega_k\hat{b}^{\dag}_{k,v}\hat{b}_{k,v}$.
The transformed system Hamiltonian is
\begin{eqnarray}~\label{hs2}
\hat{H}_s=-\epsilon_0\hat{J}_z-\xi\hat{J}^2_z,
\end{eqnarray}
where $\xi=4\sum_{k,v}\frac{\lambda^2_{k,v}}{\omega_k}$ is the reorganized energy, contributing to the collective
excitations of qubits system.
Given the Ohmic spectrum, the reorganized energy is specified as
\begin{eqnarray}
\xi&=&\frac{1}{\pi}\int^{\infty}_{0}d{\omega}\sum_{k,v}\frac{4\pi\lambda^2_{k,v}}{\omega_k}\delta(\omega-\omega_k)
=\frac{1}{\pi}\sum_v\int^{\infty}_{0}d{\omega}\frac{J_v(\omega)}{\omega}\\
&=&\frac{1}{\pi}\sum_{v}\alpha_v\omega_{cv},\nonumber
\end{eqnarray}
which is linearly proportional to the system-bath coupling strength.
The eigenbasis is  $|j,m{\rangle}$ for the Hamiltonian at Eq.~(\ref{hs2}),
and the eigenvalue is expressed as $\hat{H}_s|j,m{\rangle}=E_{j,m}|j,m{\rangle}$,
with $E_{j,m}=-\epsilon_0m-\xi{m^2},~m=-j,-j+1,...,j$.
It is interesting to note that for the single qubit case ($N=1$), the collective-qubit model is reduced to
the well-known nonequilibrium spin-boson model~\cite{ajleggett1987rmp1,dsegal2006prb1,kavelizhanin2010jcp1}.
The photonic bath mediated long-range coupling term
$-\xi{\hat{J}^2_z}$ is simplified to the constant parameter$-\xi/4$, which can be safely ignored to study the
transient dynamics and steady state behaviors.
For two or more qubits ensembles, the influence of this collective term on the dynamics becomes apparent
beyond the weak system-bath coupling regime, mainly due to the reorganization of the system energy levels $E_{j,m}$.
Whereas such feature can not be unraveled from the spin-boson model.

The transformed system-bath interaction is given by
\begin{eqnarray}~\label{vsb2}
\hat{V}_{sb}=\frac{\Delta}{2}(e^{-i\hat{B}}\hat{J}_++e^{i\hat{B}}\hat{J}_-).
\end{eqnarray}
It is clearly shown that the photon transfer process is contributed by the excitation
of the collective-qubit system from lower state to higher one with absorbing photons from the baths,
or relaxation of the qubits from higher state to the lower one by emitting photons to the baths.
Moreover, spin flips are accompanied by multi-photons collective transport.
This intrinsic picture may be difficult to uncover from the original Hamiltonian at Eq.~(\ref{ham1}).

\subsection{Quantum kinetic equation}
We apply the nonequilibrium noninteracting blip approximation (NIBA)~\cite{ajleggett1987rmp1,hdekker1987pra1,lnicolin2011prb1,lnicolin2011jcp2,tchen2013prb1} to study the collective-qubit system dynamics.
Born approximation is included to perturb the transformed Hamiltonian at Eq.~(\ref{vsb2}) up to the second order.
It is known that this approximation is valid for weak tunneling $\Delta{<}\omega_{c,v}$, at high temperature and/or strong system-bath coupling regime~\cite{ajleggett1987rmp1}, where $\omega_{c,v}$ is the cutoff frequency of the $v$th bath.
Moreover, we consider the Markov approximation, based on the assumption that the relaxation time of photonic baths is
much shorter than the counterpart of the qubits system.
Then, the quantum kinetic equation is expressed as
\begin{eqnarray}~\label{pd1}
\frac{dP_m}{dt}=-(\kappa^{-}_{m-1}+\kappa^{+}_m)P_m+\kappa^{+}_{m-1}P_{m-1}+\kappa^{-}_mP_{m+1},
\end{eqnarray}
where $P_m={\langle}j,m|\hat{\rho}_s(t)|j,m{\rangle}$ is the qubits population at angular momentum state $|j,m{\rangle}$.
The transition rates $\kappa^{\pm}_m$ excite (relax) qubits from state $|j,m{\rangle}$ to $|j,m+1{\rangle}$~(from state $|j,m+1{\rangle}$ to $|j,m{\rangle}$), demonstrating the cooperative transfer process between the qubits and photonic baths.
They are expressed as
\begin{eqnarray}~\label{rate0}
\kappa^{\pm}_{m}=(\frac{\Delta}{2})^2\frac{g^{+}_m}{2\pi}\int^{\infty}_{-\infty}d{\omega}C^{\pm}_m(\omega)d{\omega},
\end{eqnarray}
where $C^{\pm}_m(\omega)=C_S({\mp}\omega)C_D({\pm}\omega{\mp}\Delta_m)$ is the transition kernel,
and the energy gap between level $m$ and $m+1$ is $\Delta_m=E_{m+1}-E_m$.
$C_v(\omega)=\int^{\infty}_{-\infty}dte^{i{\omega}t-Q_v(t)}~(v=S,D)$ is the probability density of the $v$th bath,
describing the absorption of energy $\omega$  from the $v$the photon bath if $\omega>0$,
or the release of energy $-\omega$ if $\omega<0$.
$Q_v(t)=\sum_k(\frac{2\lambda_{k,v}}{\omega_k})^2[(2n_{k,v}+1)(1-\cos\omega_k\tau)+i\sin\omega_k\tau]$
is the photon propagator from $v$th bath, which is obtained by the thermodynamic statistics of collective photon momentums
$e^{-Q_v(t)}={\langle}e^{-i\hat{B}_v(t)}e^{i\hat{B}_v(0)}{\rangle}$.
It is found that the local $v$th transition density fulfill the detailed balance relation as $C_v(-\omega)/C_v(\omega)=e^{-\beta_v\omega}$.
However, the transition kernel $C^{\pm}(\omega)$ breaks this relation, due to the nonequilibrium bias ($T_S{\neq}T_D$).
This implies the complex transport processes, cooperatively contributed by two photon baths.
Moreover, it is found from Eq.~(\ref{rate1}) that the transition process is described by the perturbation
to the lowest order of the coherent tunneling $\Delta^2$, whereas photons are non-perturbatively involved in the energy transport process.

Since we mainly focus on the physical behaviors in the high temperature $T_{S(D)}>\epsilon_0$ and/or strong system-bath coupling regime.
The well-known Marcus limit is considered for calculating simplification~\cite{ramarcus1956jcp1}.
Briefly, it can be achieved by expanding the photon propagator $Q_v(\tau)$ in short-time regime as
$Q_v(\tau)=\xi_v(\tau^2/\beta_v+i\tau)$, with the coupling strength
$\xi_v={\alpha_v\omega_{c,v}}/{\pi}$.
Consequently, we obtain the transition kernel
$C_v(\omega)=\sqrt{\frac{\beta_v\pi}{\xi_v}}e^{-\beta_v(\omega-\xi_v)^2/4\xi_v}$,
and the transition rate
\begin{eqnarray}~\label{rate1}
\kappa^{\pm}_m=(\frac{\Delta}{2})^2g^+_m\sqrt{\frac{\pi}{T_S\xi_S+T_D\xi_D}}
\exp{[-\frac{(\Delta_m{\pm}\xi_S{\pm}\xi_D)^2}{4(T_S\xi_S+T_D\xi_D)}]}.
\end{eqnarray}
By inserting this expression of rates into the equation of motion at Eq.~(\ref{pd1}),
we are able to analyze the collective steady state behaviors with low computational cost.

Unlike the Redfield scheme in the weak coupling regime,
which shows resonant and additive transport behaviors~\cite{mvogl2011aop1,mvogl2012pra1},
population dynamics at Eq.~(\ref{pd1}) describes the nonresonant photonic energy transport process
~\cite{tchen2013prb1,lnicolin2011prb1,lnicolin2011jcp2},
which is clearly demonstrated from the transition rate at Eq.~(\ref{rate1}).
Specifically, $\kappa^{+}_m$ demonstrates that as the collective-qubit system absorbs energy $\Delta_m$ through the transition
from the state $|j,m{\rangle}$ to $|j,m+1{\rangle}$, the source bath will contribute $\omega$, and the drain bath will supply the rest $\Delta_m-\omega$.
While $\kappa^{-}_m$ means that as the qubits release energy $\Delta_m$
from the state $|j,m+1{\rangle}$ to $|j,m{\rangle}$,
the source bath will gain the $\omega$ and the drain bath will absorb the rest $\Delta_m-\omega$.
Moreover, it should be noted that a similar derivation of the quantum kinetic equation for the multi-level model was
obtained~\cite{lnicolin2011prb1}, which focused on the quantum fluctuation theorem.
In this paper, we mainly exploit the influence of the strong system-bath coupling on collective transport
of the finite-size nonequilibrium qubits system.



\begin{figure}[tbp]
\begin{center}
\includegraphics[scale=0.55]{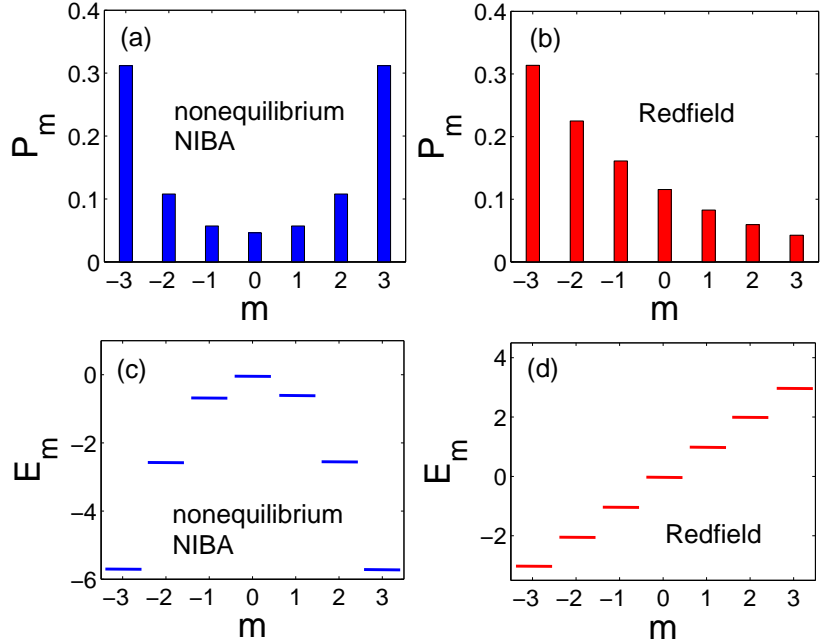}
\end{center}
\caption{(Color online) Comparisons of steady state population distribution ((a) and (b)) and eigen-energy levels ((c) and (d))
within nonequilibrium NIBA at Eq.~(\ref{pd1}) and Redfield scheme at Ref.~\cite{mvogl2011aop1}.
The parameters are given by $N=6$, $\epsilon_0=0$, $\omega_c=10\Delta$, $\alpha=0.1\Delta$,
$T_S=4\Delta$ and $T_D=2\Delta$.
}~\label{fig:fig1}
\end{figure}

\subsection{Steady state population}
We study population distributions after long time evolution $\lim_{t{\rightarrow}\infty}\frac{dP_m(t)}{dt}$
$=0$.
From the equation of motion for population dynamics at Eq.~(\ref{pd1}),
the population relation of nearest-neighboring states is given by
$P_{m+1}/P_m=\kappa^{+}_m/\kappa^{-}_m$.
For the single photonic bath, it is reduced to $P_{m+1}/P_m=e^{-\beta\Delta_m}$, due to the detailed balanced relation
of the transition processes.
While under the thermodynamic bias (i.e. $T_S>T_D$), such local balance will break.
Unlike the Redfield scheme, where the effective 'equilibrium' state can be quantified through the additive feature of photonic baths~\cite{mvogl2011aop1},
it is difficult to analytically obtain such effective distribution in strong coupling within NIBA,
due to the nonadditive transfer behavior of photonic baths.
However, for small reorganized energy $\xi{\ll}\epsilon_0$ at off-resonant case, the energy gap can be approximated by
$\Delta_m=\epsilon_0+(2m+1)\xi{\approx}\epsilon_0$, which becomes level independent.
Hence, the transition rates can be simplified to
$\kappa^{{\pm}}_m=g^{+}_m\gamma_{\pm}$,
with the rate kernel
$\gamma_{\pm}=({\Delta}/{2})^2\frac{1}{2\pi}\int^{\infty}_{-\infty}d{\omega}C_{\pm}(\omega)$
and the probability density $C_{\pm}(\omega)=C_S({\mp}\omega)C_D({\pm}\omega{\mp}\Delta)$.
To show the expression of population distribution, we assume $P_m=ay^{m}$,
and replace it to the kinetic equation at Eq.~(\ref{pd1}), resulting in $y=\gamma_+/\gamma_-$.
Moreover, through the conservation of the probability $\sum_mP_m=1$.
the constant parameter is obtained as $a=\frac{y^j(1-y)}{1-y^{2j+1}}$ with $j=N/2$.
Hence, the population distribution is shown as
\begin{eqnarray}~\label{pm1}
P_m=\frac{y^{N/2}(1-y)}{1-y^{N+1}}y^{m}.
\end{eqnarray}
It should be noted that though the profile of the steady state population at Eq.~(\ref{pm1}) within nonequilibrium NIBA  seems
the same as in the Redfield scheme~\cite{mvogl2011aop1}, the transition ratio $y$ from two schemes are quite different,
which results from the distinct transfer pictures~\cite{tchen2013prb1}.

For strong system-bath coupling, by applying the nonequilibrium NIBA approach we numerically plot the steady state population at
resonance case ($\epsilon_0=0$), as shown at Fig.~\ref{fig:fig1}(a).
It is found that population shows non-monotonic behavior in eigenbasis $|j,m{\rangle}$,
with maximal probabilities at two highest excited states $|j,{\pm}j{\rangle}$.
Moreover, it is symmetrically arranged from $m=0$.
In sharp contrast, the population distribution within the Redfield method is expressed as the monotonic decrease at Fig.~\ref{fig:fig1}(b),
with the  maximal probability expected at ground state $|j,-j{\rangle}$.
The difference of population based on two schemes mainly originates from the long-range interaction of qubits, mediated by photonic baths.
Specifically, at strong coupling regime, the influence of many photon excitations on the indirect qubit-qubit interaction becomes apparent.
This results in the nonlinear term $-\xi\hat{J}^2_z$ at Eq.~(\ref{hs2}), which is unraveled by the nonequilibrium NIBA.
Then, the energy levels of qubits system are changed to $E_m=-\xi{m^2}$ shown at Fig.~\ref{fig:fig0}(c),
with lowest levels at states $|j,{\pm}j{\rangle}$.
Considering the steady state transition balanced relation $\kappa^{+}_mP_m=\kappa^{-}_mP_{m+1}$
derived from Eq.~(\ref{pd1}), and the expressions of transition rates at Eq.~(\ref{rate1}),
we are able to obtain that two highest excited state occupy the maximal probability and
state $|j,0{\rangle}$ has the minimal probability.
However, the energy levels in Redfield scheme $E_m=\Delta{m}$ increase with angular state $m$ at Fig.~\ref{fig:fig1}(d),
which means that the steady state population should decrease accordingly.
Hence, we conclude that the indirectly collective interaction between qubits is crucial for steady state behavior in strong system-coupling regime.
Whereas it can not be exploited by the Redfield scheme, which is usually applicable in the weak coupling limit.

\section{Full counting statistics of photonic energy flux}
By applying the full counting statistics~\cite{mesposito2009rmp1,mcampisi2011rmp1}, we next obtain the expression of cumulant generating function of the nonequilibrium collective-qubit system,
to count the photon energy transport from hot source to cold drain.
By including the counting field parameter to count photon energy into the drain bath,
the Hamiltonian at Eq.~(\ref{ham2}) is changed to
$\hat{H}(\chi)=e^{i\chi\hat{H}_D/2}\hat{H}e^{-i\chi\hat{H}_D/2}$,
with $\hat{H}_D=\sum_k\omega_k\hat{b}^{\dag}_{k,D}\hat{b}_{k,D}$ the drain bath Hamiltonian.
Then, under the modified canonical transformation, with the counting parameter embedded unitary operator
$\hat{P}_{\chi}=e^{i\hat{J}_z\hat{B}_{\chi}}$ and the modified collective bath momentum
$\hat{B}_{\chi}=i\sum_{k,v}\frac{2\lambda_{k,v}}{\omega_k}e^{i\chi\omega_k\delta_{v,D}/2}\hat{b}^{\dag}_{k,v}+H.c.$,
the Hamiltonian is modified to $\hat{H}_t(\chi)=\hat{H}_s+\hat{H}_b+\hat{V}_{sb}(\chi)$.
The reorganized system-bath interaction is described as
$\hat{V}_{sb}(\chi)=\frac{\Delta}{2}(e^{-i\hat{B}_{\chi}}\hat{J}_++e^{i\hat{B}_{\chi}}\hat{J}_-)$.
Similar to the procedures to obtain the Eq.~(\ref{pd1}) under the Markovian-NIBA approximation,
the quantum kinetic equation combined with the counting field parameter is given by
\begin{eqnarray}~\label{pd2}
\frac{dP^{\chi}_m}{dt}&=&-(\kappa^-_{m-1}+\kappa^+_m)P^{\chi}_m
+\kappa^{+}_{m-1}(\chi)P^{\chi}_{m-1}+\kappa^{-}_{m}(\chi)P^{\chi}_{m+1},
\end{eqnarray}
where the modified transition rates are
\begin{eqnarray}~\label{rate2}
\kappa^{\pm}_m(\chi)=(\frac{\Delta}{2})^2\frac{g^+_m}{2\pi}e^{{\mp}i\Delta_m\chi}
\int^{\infty}_{-\infty}d{\omega}C^{\pm}_m(\omega)e^{{\pm}i\omega\chi}d\omega.
\end{eqnarray}
In absence of the counting field parameter, $P^{\chi=0}_m$ are the state populations $P_m$,
and Eq.~(\ref{pd2}) is reduced
to the standard quantum kinetic equation at Eq.~(\ref{pd1}).
Meanwhile, the modified transition rates return back to Eq.~(\ref{rate0}).
In the Marcus limit, these transition rates at Eq.~(\ref{rate2}) are simplified to
$\kappa^{\pm}_m(\chi)=\kappa^{\pm}_{m}F^{\pm}_m(\chi)$, with $\kappa^{\pm}_m$ shown at Eq.~(\ref{rate1})
and the counting factor
\begin{eqnarray}
F^{\pm}_m(\chi)=\exp\{{\mp}i\Delta_m\chi-\frac{T_ST_D\xi_S\xi_D}{T_S\xi_S+T_D\xi_D}
[i\chi(\frac{1}{T_S}+\frac{\mp\Delta_m-\xi_D}{T_D\xi_D})+\chi^2]\}.
\end{eqnarray}

Then by re-arranging populations in column style
$|\mathrm{P}(\chi,t){\rangle}=[P^{\chi}_{-j},P^{\chi}_{-j+1},...,P^{\chi}_{j}]^{T}$, Eq.~(\ref{pd2}) can
be re-expressed as $\frac{\partial}{\partial{t}}|\mathrm{P}(\chi,t){\rangle}=-\mathcal{L}(\chi)|\mathrm{P}(\chi,t){\rangle}$, with
$\mathcal{L}(\chi)$ tri-diagonal matrix dependent on the transition rates.
Hence, the generating function can be formally obtained as
$G(\chi)=\lim_{t{\rightarrow}\infty}\frac{1}{t}\ln{\langle}I|\mathrm{P}(\chi,t){\rangle}$,
with the left unit vector ${\langle}I|=[1,1,...,1]$.
For the long-time limit behaviors are of primer interest in this study,
the corresponding cumulant generating function is practically approaching the ground state energy $E_0(\chi)$ of transfer matrix $\mathcal{L}(\chi)$,
having the smallest real part.
Based on the cumulant generating function, the $n$th order cumulant of photonic energy current fluctuations can be steadily
obtained as $\lim_{t{\rightarrow}\infty}{\langle}{\langle}\hat{Q^{n}}{\rangle}{\rangle}/t
=\frac{{\partial}^nG(\chi)}{\partial{(i\chi)}^{n}}|_{\chi=0}$, and the steady state energy flux is the lowest order
$J=\frac{{\partial}G(\chi)}{\partial{(i\chi)}}|_{\chi=0}$.

\begin{figure}[tbp]
\begin{center}
\includegraphics[scale=0.6]{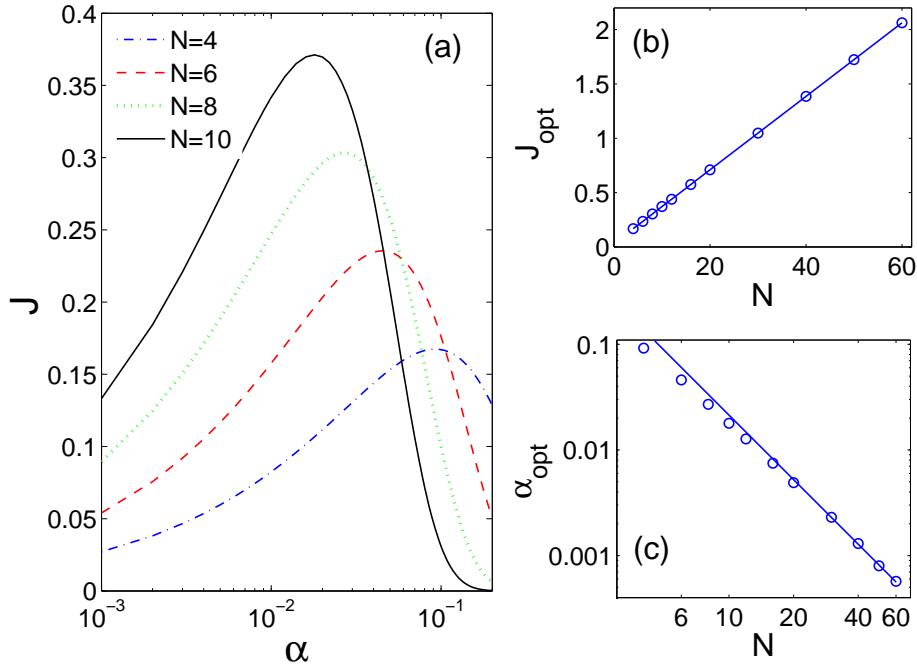}
\end{center}
\caption{(Color online) (a) Photonic energy flux of finite-size nonequilibrium qubits system in wide system-bath coupling regime;
(b) finite-size scaling of the optimal energy flux:$J_{opt}=\max_{\alpha}\{J\}$;
(c) finite-size scaling of the optimal system-bath coupling, which corresponds to the optimal flux.
The parameters are given by $\epsilon_0=0$, $\omega_c=10\Delta$, $T_S=4\Delta$ and $T_D=2\Delta$.
}~\label{fig:fig3}
\end{figure}

\begin{figure}[tbp]
\begin{center}
\includegraphics[scale=0.45]{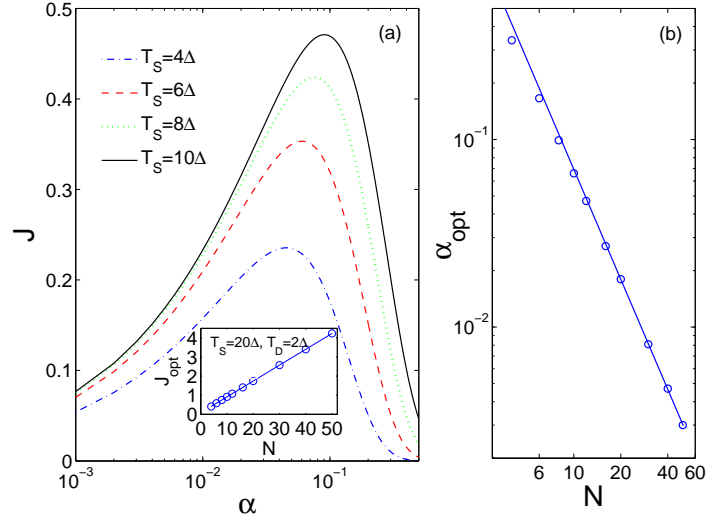}
\end{center}
\caption{(Color online) (a) Photonic energy flux of finite-size nonequilibrium qubits system with different thermodynamic biases;
(b) finite size scaling of the optimal flux:$J_{opt}=\max_{\alpha}\{J\}$ with bath temperatures $T_S=20\Delta$ and $T_D=2\Delta$;
(c) finite size scaling of the optimal system-bath coupling with bath temperatures $T_S=20\Delta$ and $T_D=2\Delta$, which corresponds to the optimal flux.
The other parameters are given by $\epsilon_0=0$ and $\omega_c=10\Delta$.
}~\label{fig:fig4}
\end{figure}

\subsection{Steady state energy flux}
We study the influence of system-bath coupling on the steady state photonic energy flux beyond the weak interaction regime,
as shown at Fig.~\ref{fig:fig3}(a).
By increasing coupling strength, the energy flux is firstly raised to reach the summit.
It mainly results from the contribution of the nonresonant photon transfer processes.
If we further enlarge the interaction strength to the ultra-strong coupling regime, energy flux is suppressed
due to the detrimental photon-qubit scattering.
As a result, the non-monotonic feature appears, which provides guide to the optimal control of the photonic flux.
This optimal behavior is similar to the nonequilibrium spin-boson model within NIBA scheme~\cite{lnicolin2011jcp2,tchen2013prb1}, which is the special case of the collective-qubit model by setting $N=1$.
Next, we analyze the finite-size effect of the energy flux.
By increasing the system size, the optimal value of the energy flux is accordingly amplified.
Specifically, the optimal flux shows linear scaling with the system size $J_{opt}{\propto}N$, shown at Fig.~\ref{fig:fig3}(b).
Moreover, the optimal coupling strength shrinks with the increase of the system size at Fig.~\ref{fig:fig3}(c),
exhibiting inversed power-law scaling $\alpha_{opt}{\propto}N^{-\gamma}$, with the scaling exponent $\gamma=2.00{\pm}0.06$.
It should be noted that these novel features can not be observed within the Redfield scheme~\cite{mvogl2011aop1},
where the resonant energy flux linearly increases by enlarging the system-bath coupling,
and no optimal behavior emerges.

Then, we turn to study the finite-size scaling effect of the energy flux under high temperature bias
(\emph{e.g.}, $T_S=20\Delta,~T_D=2\Delta$) to find the universal feature, shown at Fig.~\ref{fig:fig4}.
It was reported that within Redfield weak-coupling limit, the energy flux in the high temperature regime
asymptotically  reaches summit, and superradiant transport signal is found($J{\propto}N^2$)~\cite{mvogl2011aop1}.
However, in  strong coupling regime, the energy flux exhibits nonmonotonic behavior,
which demonstrates disappearance of the peak platform in high temperature bias.
Besides, its optimal value shows the linear scaling with the system size,
which implies that the linear form may be universal for the optimal photonic flux.
This results seem to be a sharp contrast to the counterpart in the weak coupling regime.
However, the deviations in the same quantum system between two schemes just reflect the different physical manifestations.
Hence, they are complimentary to each other to fully unravel novel system features.
Moreover, the coupling strength corresponding to the optimal energy flux shows inversed power-law case
$\alpha_{opt}{\propto}N^{-\gamma},~\gamma=2.00{\pm}0.10$.
This strengthens the proposal that the finite-size scaling of the optimal coupling strength may be universal.

\begin{figure}[tbp]
\begin{center}
\includegraphics[scale=0.6]{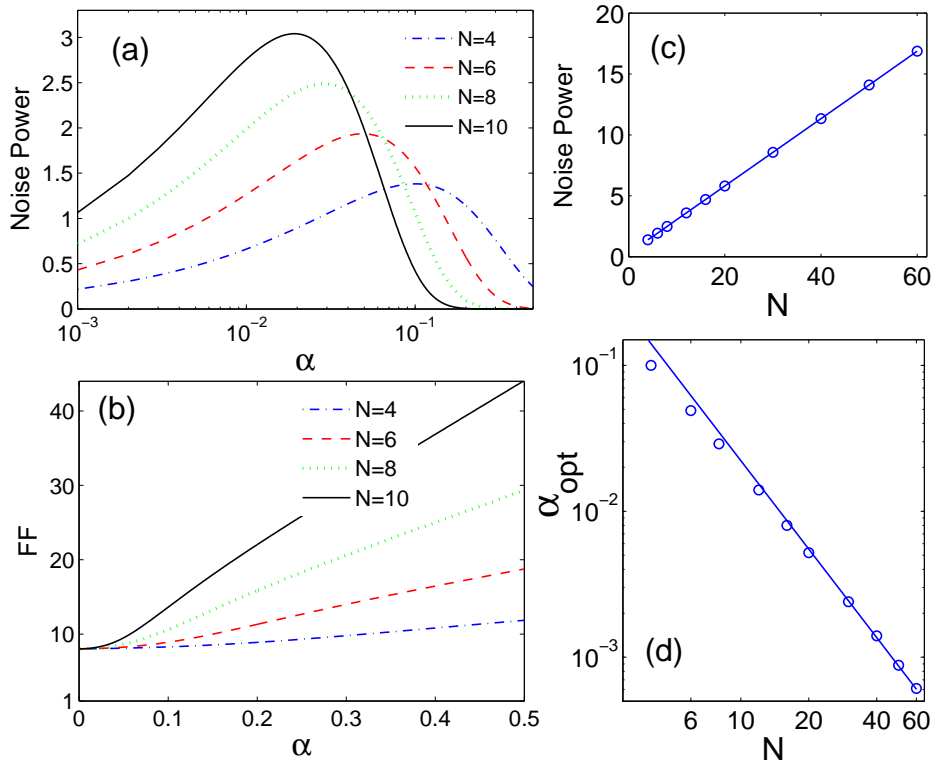}
\end{center}
\caption{(Color online) (a) Steady state behaviors of noise power with different system sizes;
(b) ratio of the noise power to photonic energy flux;
(c) finite-size scaling of the optimal noise power;
(d) finite-size scaling of the optimal coupling strength, corresponding to the optimal noise power.
The parameters are given by $\epsilon_0=0$, $\omega_c=10\Delta$, $T_S=4\Delta$ and $T_D=2\Delta$.
}~\label{fig:fig5}
\end{figure}

\subsection{Noise power}
We study the influence of  system-bath interaction on the noise power,
defined by $S={{\partial}^2G(\chi)}/{{\partial}(i\chi)^2}|_{\chi=0}$.
By tuning the coupling strength into strong regime, the monotonic behavior of the noise power is clearly exhibited given by Fig.~\ref{fig:fig4}(a), which is similar to the photonic energy flux at Fig.~\ref{fig:fig3}(a).
To further describe the statistic feature of fluctuation of the energy flux,
we introduce $FF=({\langle}\hat{Q}^2{\rangle}-{\langle}\hat{Q}{\rangle}^2)/{\langle}\hat{Q}{\rangle}=
S/J$ with $\hat{Q}=\sum_k\omega_k\hat{b}^{\dag}_{k,D}\hat{b}_{k,D}$, defined as the ratio of the standard deviation
to the mean.
In the weak coupling regime, it shows a platform, which is consistent to the counterpart within the Redfield scheme,
independent of the coupling strength~\cite{mvogl2011aop1}.
Interestingly, it shows almost linear behavior by enlarging the coupling strength, shown at Fig.~\ref{fig:fig5}(b).
This demonstrates that the power noise is more sensitive than the energy flux in strong coupling regime
as a measuring indicator.
Moreover, we study the finite-size scaling of the optimal noise power at Fig.~\ref{fig:fig5}(c),
which shows the linear feature.
Hence, the noise signal can be amplified by increasing the qubits system size.
Then, we analyze the relation of optimal coupling strength with the system size, given by Fig.~\ref{fig:fig5}(c).
The universal scaling feature is unraveled as $\alpha_{opt}{\propto}N^{-\gamma},~\gamma=2.00{\pm}0.10$.
Though the higher orders (e.g., skewness) are not shown, the finite-size scaling effects of the optimal value and the optimal coupling strength are the same as those in the noise power.
As a result, we conclude that both the linear scaling behavior of the optimal flux fluctuations and
the power-law scaling behavior of the optimal coupling strength may be universal for the collective-qubit system in strong coupling regime.


\section{Conclusion}
In summary, the nonequilibrium NIBA approach is applied to study the steady state quantum transport of collective-qubit system in strong
photon-qubit coupling regime.
The collective term of indirect qubit-qubit interaction, emerging from the cooperative contribution of many photon excitations, is essential to establish the novel steady
state distribution, which is apparently deviated from the counterpart within Redfield scheme.
The turnover behavior of photonic energy flux and high order cumulants (e.g., noise power) is analyzed for finite-size qubits system, which implies the optimal control of the collective quantum transport.
Moreover, the linear scaling feature of both optimal energy flux and high order cumulants is described, even in high
temperature bias limit.
High order cumulants show more sensitive signatures than the photonic energy flux by tuning coupling strength into strong regime.
Interestingly, the inversed power-law scaling feature of the optimal coupling strength is unraveled, which is suggested to be universal for the nonequilibrium collective-qubit system.
It is known that to detect nonmonotonic behavior of spin-boson system is still a challenging problem~\cite{kavelizhanin2010jcp1,ksaito2013prl1}.
Here, with such scaling effect, we believe it may provide an alternative road to measuring such nonmonotonic behavior of spin-boson system in comparatively weak coupling regime, by only increasing the qubits number.

\section{acknowledgement}
We acknowledge support by the Program for Innovative Research Team in Hangzhou Dianzi University.
K-W. Sun is supported by National Natural Science Foundation of China (Grant No. 11404084).

\appendix
\widetext
\section*{Appendix: Quantum master equation with full counting statistics}
The general expression of second-order quantum master equation under Born-Markov approximation is given by
\begin{eqnarray}
\frac{d\hat{\rho}^{\chi}_s}{dt}&=&-i[\hat{H}_s,\hat{\rho}^{\chi}_s]
-\int^{\infty}_0d\tau\textrm{Tr}_b\{[\hat{V}_{\chi},[\hat{V}_{\chi}(-\tau),\hat{\rho}^{\chi}_s]_{\chi}]_{\chi}\},
\end{eqnarray}
where $\hat{\rho}^{\chi}_s$ is the modified reduced system density matrix, by tracing the photonic bath degrees off the total density matrix,
and $[\hat{A}_{\chi},\hat{B}_{\chi}]_{\chi}=\hat{A}_{\chi}\hat{B}_{\chi}-\hat{B}_{\chi}\hat{A}_{-\chi}$.
Starting from the reorganized system $\hat{H}_s=\epsilon_0\hat{J}_z-\xi\hat{J}^{2}_{z}$ at Eq.~(\ref{hs2}),
and the modified system-bath interaction
$\hat{V}_{sb}(\chi)=\frac{\Delta}{2}(e^{-i\hat{B}_{\chi}}\hat{J}_{+}+e^{i\hat{B}_{\chi}}\hat{J}_{-})$
with the collective photonic momentum
$\hat{B}_{\chi}=i\sum_{k,v}\frac{2\lambda_{k,v}}{\omega_k}(e^{i\omega_k\chi\delta_{v,D}/2}\hat{b}^{\dag}_{k,v}-H.c.)$,
the population dynamics under angular basis $\{|j,m{\rangle}\}$ with the counting field parameter is given by
\begin{eqnarray}
\frac{dP^{\chi}_m}{dt}&=&-(\frac{\Delta}{2})^2\int^{\infty}_0d{\tau}\{[C(\tau)e^{i\Delta_{m-1}\tau}g^{+}_{m-1}P^{\chi}_m
+C(\tau)e^{-i\Delta_m\tau}g^{+}_mP^{\chi}_m]\\
&&+[C^{*}(\tau)e^{-i\Delta_{m-1}\tau}g^{+}_{m-1}P^{\chi}_m
+C^{*}(\tau)e^{i\Delta_m\tau}g^{+}_mP^{\chi}_m]\nonumber\\
&&-[C^{*}(\tau+\chi)e^{i\Delta_{m-1}\tau}g^{+}_{m-1}P^{\chi}_{m-1}+C^{*}(\tau+\chi)e^{-i\Delta_m\tau}g^{+}_mP^{\chi}_{m+1}]\nonumber\\
&&-[C(\tau-\chi)e^{-i\Delta_{m-1}\tau}g^{+}_{m-1}P^{\chi}_{m-1}+C(\tau-\chi)e^{i\Delta_m\tau}g^{+}_mP^{\chi}_{m+1}]\}\nonumber
\end{eqnarray}
where the time domain correlation function is
\begin{eqnarray}
C(\tau)={\langle}e^{-i\hat{B}(\tau)}e^{i\hat{B}}{\rangle}_b=\exp\{-\sum_{k,v}(\frac{2\lambda_{k,v}}{\omega_k})^2
[(2n_k+1)(1-\cos\omega_k\tau)+i\sin\omega_k\tau]\},\nonumber
\end{eqnarray}
the angular coefficient is $g^{+}_{m}=j(j+1)-m(m+1)$ with $j=N/2$,
and energy gap is $\Delta_m=E_{m+1}-E_m=\epsilon_0-(2m+1)\xi$ with  eigenvalue $E_m=(\epsilon_0m-\xi{m^2})$.
By defining the transition rates
\begin{eqnarray}
\kappa^{\pm}_m(\chi)
=(\frac{\Delta}{2})^2g^+_m\int^{\infty}_{-\infty}d{\tau}
C(\tau-\chi)e^{{\mp}i\Delta_{m}\tau},\nonumber
\end{eqnarray}
the equation of motion for population dynamics is simplified as
\begin{eqnarray}
\frac{dP^{\chi}_m}{dt}&=&-(\kappa^-_{m-1}+\kappa^+_m)P^{\chi}_m
+\kappa^{+}_{m-1}(\chi)P^{\chi}_{m-1}+\kappa^{-}_{m}(\chi)P^{\chi}_{m+1},
\end{eqnarray}
with $\kappa^{\pm}_m=\kappa^{\pm}_m(\chi=0)$.
In absence of the counting field, the modified equation of motion is returned to the standard quantum kinetic equation
for the system populations as $P^{\chi=0}_m=P_m$.


\begin{thebibliography}{99}
\bibitem{awallraff2004nature1} A. Wallraff, D. I. Schuster, A. Blais, L. Frunzio, R.-S. Huang, J. Majer, S. Kumar,
S. M. Girvin, and R. J. Schoelkopf, Nature (London) \textbf{431}, 162 (2004).
\bibitem{tbrandes2005physrep1} T. Brandes, Phys. Rep. \textbf{408}, 315 (2005).
\bibitem{lzhou2008prl1} L. Zhou, Z. R. Gong, Y. X. Liu, C. P. Sun, and F. Nori, Phys. Rev. Lett. \textbf{101},
100501 (2008).
\bibitem{dzxu2013pra1} D. Z. Xu, Y. Li, C. P. Sun, and P. Zhang, Phys. Rev. A \textbf{88}, 013832 (2013).
\bibitem{dwwang2015prl1} D. W. Wang, R. B. Liu, S. Y. Zhu, and M. O. Scully, Phys. Rev. Lett. \textbf{114}, 043602 (2015).
\bibitem{jklinder2015pnas1} J. Klinder, H. Kebler, M. Wolke, L. Mathey, and A. Hemmerich, PNAS \textbf{112}, 3290 (2015).
\bibitem{mfeng2015nc1} M. Feng, Y. P. Zhong, T. Liu, L. L. Yan, W. L. Yang, J. Twamley, and H. Wang, Nat. Comm. \textbf{6}, 7111 (2015).
\bibitem{mmohseni1} M. Mohseni and Y. Omar, \emph{Quantum Effects in Biology} (Cambridge University Press, 2014).
\bibitem{hritsch2013rmp1} H. Ritsch, P. Domokos, F. Brennecke, and T. Esslinger, Rev. Mod. Phys. \textbf{85}, 553 (2013).
\bibitem{krahazzard2014prl1} K. R. A. Hazzard, B. Gadway, M. Foss-Feig, B. Yan, S. A. Moses, J. P. Covey, N. Y. Yao,
M. D. Lukin, J. Ye, D. S. Jin, and A. M. Rey, Phys. Rev. Lett. \textbf{113}, 195302 (2014).
\bibitem{rhdicke1954pr1} R. H. Dicke, Phys. Rev. \textbf{93}, 99 (1954).
\bibitem{nlambert2004prl1} N. Lambert, C. Emary, and T. Brandes, Phys. Rev. Lett. \textbf{92}, 073602 (2004).
\bibitem{qhchen2008pra1} Q. H. Chen, Y. Y. Zhang, T. Liu, and K. L. Wang, Phys. Rev. A \textbf{78}, 051801 (2008).
\bibitem{tvorrath2004cp1} T. Vorrath and T. Brandes, Chem. Phys. \textbf{296}, 295 (2004).
\bibitem{tvorrath2004prl1} T. Vorrath and T. Brandes, Phys. Rev. Lett. \textbf{95}, 070402 (2005).
\bibitem{fbanders2008njp1} F. B. Anders, New J. Phys. \textbf{10}, 115007 (2008).
\bibitem{mesposito2009rmp1} M. Esposito, U. Harbola, and S. Mukamel, Rev. Mod. Phys. \textbf{81}, 1665 (2009).
\bibitem{mcampisi2011rmp1} M. Campisi, P. H\"{a}nggi, and P. Talker, Rev. Mod. Phys. \textbf{83}, 771 (2011).
\bibitem{mvogl2011aop1} M. Vogl, G. Schaller, and T. Brandes, Annal of Phys. \textbf{326}, 2827 (2011).
\bibitem{mvogl2012pra1} M. Volgl, G. Schaller, E. Sch\"{o}ll, and T. Brandes, Phys. Rev. A \textbf{86}, 033820 (2012).
\bibitem{ajleggett1987rmp1} A. J. Leggett, Rev. Mod. Phys. \textbf{59}, 1 (1987).
\bibitem{dsegal2006prb1} D. Segal, Phys. Rev. B \textbf{73}, 205415 (2006).
\bibitem{kavelizhanin2010jcp1} K. A. Velizhanin, H. Wang, and M. Thoss, J. Chem. Phys. \textbf{133}, 084503 (2010).
\bibitem{tchen2013prb1} T. Chen, X. B. Wang, and J. Ren, Phys. Rev. B \textbf{87}, 144303 (2013).
\bibitem{cwang2015sr1} C. Wang, J. Ren, and J. S. Cao, Sci. Rep. \textbf{5}, 11787 (2015).
\bibitem{hdekker1987pra1} H. Dekker, Phys. Rev. A \textbf{35}, 1436 (1987).
\bibitem{lnicolin2011prb1} L. Nicolin and D. Segal, Phys. Rev. B \textbf{84}, 161414 (2011).
\bibitem{lnicolin2011jcp2} L. Nicolin and D. Segal, J. Chem. Phys. \textbf{135}, 164106  (2011).

\bibitem{dsegal2014pre1} D. Segal, Phys. Rev. E \textbf{90}, 012148 (2014).
\bibitem{ksaito2013prl1} K. Saito and T. Kato, Phys. Rev. Lett. \textbf{111}, 214301 (2013).
\bibitem{dsegal2015arxiv2} E. Taylor and D. Segal, Phys. Rev. Lett. \textbf{114}, 220401 (2015).

\bibitem{ramarcus1956jcp1} R. A. Marcus, J. Chem. Phys. \textbf{24}, 966 (1956).


\end{thebibliography}
\end{document}